\documentstyle[prl,aps]{revtex}
\draft
\begin{document}
\title{Quantum Self-Induced Transparency in Frequency Gap Media}
\author{Sajeev John and Valery I. Rupasov\cite{pa}}
\address{Department of Physics, University of Toronto,
Toronto, Ontario, Canada M5S 1A7}
\date{August 20, 1997}
\maketitle
\begin{abstract}
We study quantum effects of light propagation through an
extended absorbing system of two-level atoms placed within
a frequency gap medium (FGM). Apart from ordinary solitons
and single particle impurity band states, the many-particle
spectrum of the system contains massive pairs of confined
gap excitations and their bound complexes - gap solitons.
In addition, ``composite'' solitons are predicted as bound
states of ordinary and gap solitons. Quantum gap and composite
solitons propagate without dissipation, and should be associated
with self-induced transparency pulses in a FGM.
\end{abstract}

The self-induced transparency (SIT) pulses, predicted and
observed in the pioneering work of McCall and Hahn \cite{MH},
may be regarded as solitons of the Maxwell-Bloch model \cite{L},
describing classical radiation propagating in a single
direction and coupled to an extended system of two-level
atoms. The model is completely integrable \cite{AKNS} and
the time evolution of an arbitrary radiation incident on
an atomic system is described \cite{K,SIT} by the inverse
scattering method \cite{ISM}.

In the case of a high intensity pulse in ordinary vacuum,
quantum corrections are negligibly small. Therefore the
quantum version of the classical model - quantum Maxwell-Bloch
(QMB) model - has been studied \cite{QMB} only in the
context of the superfluorescence phenomenon where quantum
effects play a crucial role \cite{LC}. But the situation
is drastically changed for frequency gap media (FGM), such
as a frequency dispersive medium \cite{LL}, a photonic
bandgap material \cite{Y,J}, and a one-dimensional Bragg
reflector \cite{SS}, where classical, linear wave propagation
inside a frequency gap is excluded \cite{RS,JR}.

In this Letter, we demonstrate the existence of nonclassical
light propagation through an extended homogeneous \cite{MKK}
system of two-level atoms placed within a FGM. These light
pulses are highly correlated quantum many-body states and
are distinct from single photon hopping conductivity \cite{JQ}
through the photonic impurity band created by the atoms.
Because of a nonlocal polariton-atom coupling, an extension
of the Bethe ansatz method \cite{BA} from the case of a
single atom embedded in FGM \cite{RS,JR} to the case of
an extended many-atom system requires a thorough analysis.
The QMB generalized to the case of FGM exhibits hidden
integrability \cite{RS}, provided that the characteristic
times of interatomic resonance dipole-dipole interaction (RDDI)
and other collisional dephasing effects are much longer than
the light pulse duration. Integrability of the model allows
us to describe the time evolution of an arbitrary light pulse
incident on the system in terms of the allowed soliton modes.
Making use of the Bethe ansatz technique, we derive the Bethe
ansatz equations (BAE), which completely determine the spectrum
of the radiation plus medium plus atoms system. Here we consider
the case of the atomic transition frequency $\omega_{12}$ lying
deep inside a frequency gap of a frequency dispersive medium,
for which the McCall-Hahn theory is inapplicable.

Unlike an attractive effective photon-photon coupling
in empty space (or nondispersive media) \cite{QMB} caused
by scattering of photons on an atomic system, an effective
polariton-polariton coupling in FGM is shown to be attractive
only for polaritons of the lower polariton branch. Therefore
bound many-polariton complexes (ordinary solitons) can be
constructed only from polaritons of the lower branch. In
the limit of a macroscopically large number of polaritons,
these quantum complexes are nothing but SIT pulses
(2$\pi$-pulses) of the classical theory slightly modified
due to a nonlinear polariton dispersion.

Due to the existence of a frequency gap, the multiparticle
spectrum of the system, apart from polaritons and ordinary
solitons, also contains massive pairs of confined gap
excitations, which do not exist out of pairs, and bound
complexes of these pairs - quantum gap solitons. The
energy-momentum dispersion relations for gap solitons are
derived and the widths of soliton bands, the soliton masses,
the spatial sizes and the velocities of propagation inside
the atomic system are evaluated as functions of the number
of pairs and the atomic density. We predict also the existence
of ``composite'' solitons as bound states of ``deformed''
ordinary and gap solitons.

In contrast to quantum gap solitons generated by a single
atom, which propagate along a radial coordinate centered
at the atom \cite{JR}, quantum SIT pulses in a doped FGM
propagate in a direction defined by a single wave vector.
Furthermore, the gap SIT pulse, consisting of an even number
of gap excitations, is distinct from (odd photon number) gap
soliton hopping conduction inside the RDDI mediated impurity
band.

In the dipole, rotating wave approximation \cite{SIT}
the Hamiltonian of the generalized QMB model can be
written as $\hat{H}=\hat{H}_0+\hat{V}$, where
\begin{mathletters}
\begin{equation}
\hat{H}_0=\omega_{12}\sum_{a=1}^{M}\left(\sigma^z_a+\frac{1}{2}\right)
+\int_{C}\frac{{\rm d}\omega}{2\pi}\,\omega\,p^\dagger(\omega)p(\omega)
\end{equation}
represents the Hamiltonians of $M$ identical two-level atoms
and free polaritons, while the operator
\begin{equation}
\hat{V}=-\sqrt{\gamma}\sum_{a=1}^{M}\int_{C}\frac{{\rm d}\omega}{2\pi}
\sqrt{z(\omega)}[\sigma^+_a p(\omega){\rm e}^{ik(\omega)x_a}+
{\rm h.c.}]
\end{equation}
describes their coupling. The polariton operators $p(\omega)$
obey the commutator
$[p(\omega),p^\dagger(\omega')]=2\pi\delta(\omega-\omega')$,
while the spin operators
$\vec{\sigma}_a=(\sigma^x_a,\sigma^y_a,\sigma^z_a)$,
$\sigma^{\pm}=\sigma^x\pm i\sigma^y$ describe atoms having
the coordinates $\{x_a,\,a=1,\ldots,M\}$ on the polariton
propagation axis (the $x$ axis). The states between frequencies
labeled as $\Omega_{\bot}$ and $\Omega_{\|}$ are forbidden
for linear propagating polariton modes. Therefore, the
integration contour $C$ consists of two allowed intervals,
$C=C_-+C_+$, where $C_-=(0,\Omega_{\bot})$ and $C_+=(\Omega_{\|},\infty)$.
The coupling constant $\gamma=2\pi\omega_{12}d^2/S_0$,
where $d$ is the atomic dipole moment and $S_0$ is the
cross-section of a light beam. The information about the
medium spectrum is contained in the dispersion relation
$k(\omega)=\omega n(\omega)$. The atomic form factor
$z(\omega)=\omega n^3(\omega)/\omega_{12}$, where
$n(\omega)=\sqrt{\varepsilon(\omega)}$ and
$\varepsilon(\omega)=(\omega^2-\Omega^2_\|)/(\omega^2-\Omega^2_\bot)$
is the dielectric permeability of a frequency dispersive medium.

The eigenvalues of the model (1) are found from the following
Bethe ansatz equations
\end{mathletters}
\begin{mathletters}
\begin{equation}
\exp{(ik_jL)}\left(\frac{h_j-i\beta/2}{h_j+i\beta/2}\right)^M=
-\prod_{l=1}^{N}\frac{h_j-h_l-i\beta}{h_j-h_l+i\beta},
\end{equation}
where $k_j\equiv k(\omega_j)$, $E=\sum_{j}\omega_j$ is the
eigenenergy and the ``rapidity'' $h_j\equiv h(\omega_j)$ is given
by
\begin{equation}
h(\omega)=\frac{\omega-\omega_{12}}{\omega n^3(\omega)}.
\end{equation}
BAE have a clear physical meaning: the first phase factor
on l. h. s. is acquired by a polariton wave function during
free propagation between points $\mp L/2$, while the second
one accounts for phase factors resulting from subsequent
scattering on $M$ atoms. Propagating between the points
$\mp L/2$ the polariton is also scattered by the other $N-1$ polaritons,
and its wave function acquires the phase factor given on r. h. s.
in eq. (2a). Information concerning the nonlinear polariton
dispersion is contained in the rapidity $h(\omega)$. In empty
space, it is reasonable to neglect the resonance dipole-dipole
interaction between atoms in self-induced transparency, since
the dominant photon-atom interaction is pointlike and the
excited atom decays by stimulated emission into optical pulse
modes. In this case, light scattering from each of the $M$
atoms is considered independently. In the FGM, there are no
classical modes available for stimulated emission and the
polariton-atom coupling is highly nonlocal: An excited atom
(photon-atom bound state) exhibits nonlocal interaction
with other atoms which are within the classical tunneling
distance. This leads to coherent hopping conduction of a
photon through the resulting impurity band \cite{JQ}.
If the characteristic time of the interatomic hopping is
much longer than the pulse duration, RDDI mediated transfer
of energy between impurity atoms in an {\em arbitrary}
direction can be neglected. Energy transfer occurs through
the soliton band \cite{JR} rather than the impurity band
\cite{JQ}. Eq. (2a) is obtained by including RDDI contributions
only from virtual polaritons traveling in the {\em same}
direction as the incident pulse. In this generalization of
the single-atom soliton band to an $M$-atom soliton band,
scattering from each atom is treated independently. This
is equivalent to the independent atom (gas) approximation
used by McCall and Hahn in ordinary vacuum. Accordingly,
in ordinary vacuum [$n(\omega)=1$], eqs. (2) reduce to the
BAE of the QMB model which, for large $N$, describes the
classical McCall-Hahn solution.

It is instructive to derive the main results of standard SIT
theory from eqs. (2) with the rapidity
$h(\omega)\simeq(\omega-\omega_{12})/\omega_{12}$
corresponding to the case of empty space. As $L\to\infty$,
eqs. (2) admit solutions in which complex rapidities $h_j$
are grouped into ``strings'' containing $n$ particles,
\end{mathletters}
\begin{equation}
h_j=H+i(\beta/2)(n+1-2j),\;\;\;j=1,\ldots,n,
\end{equation}
where $H$ is a common real part (``carrying'' rapidity).
Due to the linear relationships between the rapidity,
frequency and momentum, particle frequencies and momenta
are also grouped into string structures,
$k_j=K+i(\gamma/2)(n+1-2j)$, $\omega_j=\Omega+i(\gamma/2)(n+1-2j)$,
where $K$ and $\Omega$ are common real parts of momenta
and frequencies, respectively. To avoid confusion, we use
the term ``string'' for solutions of BAE in the $h$-space and
the term ``soliton'' to refer to string's images in the
$\omega$- and $k$-spaces. Consider for simplicity the case
when all $N$ particles are grouped into a string,
i. e. $N=n$. Inserting $k_j$ and $\omega_j$ in eq. (2a)
and evaluating the product over $j=1,\ldots,N$, we obtain
the simple equation $\exp{(iQnL)}=1$, where
\begin{mathletters}
\begin{equation}
Q(\Omega)=K-\frac{2\rho}{n}\arctan{\frac{n\gamma}{2(\Omega-\omega_{12})}}
\end{equation}
and the number of atoms is represented as $M=\rho L$. Here
$\rho$ is the linear density of the number of atoms. Clearly
$Q(\Omega)$ can be interpreted as the energy-momentum
dispersion relation of a soliton of size $n$, where the second
term describes a contribution of photon-atom scattering. The
group velocity of soliton propagation $V=d\Omega/dQ$ is then
given by
\begin{equation}
\frac{1}{V}=\frac{1}{c}+
\frac{\gamma\rho}{(\Omega-\omega_{12})^2+(n\gamma/2)^2},
\end{equation}
where $c$ is the speed of light in empty space. Eq. (4b) is
identical to the corresponding expression in the classical
SIT theory. The spatial size of a soliton, $l_s\simeq(\gamma n)^{-1}$,
is inversely proportional to the number of photons \cite{QMB}.
Therefore, only macroscopically ``long'' strings, $n\gg 1$,
propagate in an absorbing atomic system without dissipation.

In FGM, eq. (3) is a solution of BAE if and only if the
imaginary parts of rapidities $h_j$ and corresponding momenta
$k_j$ have the same sign,
\end{mathletters}
\begin{equation}
{\rm sgn}({\rm Im}\,h_j)={\rm sgn}({\rm Im}\,k_j),\;\;\;j=1,\ldots,n.
\end{equation}
It is easy to understand that the necessary condition (NC)
(5) determines the frequency intervals, in which an effective
particle-particle coupling is attractive, and hence admits
bound many-particle complexes.

We start with the case when the real part of $\omega_j$
lies outside the gap. Let $\omega=\xi+i\eta$ and $\xi\in C$.
Making use of the approach developed in \cite{JR}, it is
easy to show that the effective coupling is attractive
only between polaritons of the lower branch, $\xi\in C_-$.
Polaritons of the upper branch are described by one-particle
strings with real positive rapidities and do not form any bound
complexes. Bound many-polariton complexes (ordinary solitons)
are quite similar to solitons of the QMB model, despite
their inordinate behavior on different polariton branches.
The dispersion relation of an ordinary soliton of size $n$
is given by
\begin{mathletters}
\begin{equation}
q(\xi)=k(\xi)-\frac{2\rho}{n}\arctan{\frac{\beta n}{2h(\xi)}},
\end{equation}
where $k(\xi)=\xi n(\xi)$. The group velocities inside, $V=d\xi/dq$,
and outside, $v=d\xi/dk$, the atomic system are then related by
\begin{equation}
\frac{1}{V}=\frac{1}{v}+\frac{\rho\beta}{h^2(\xi)+(\beta n/2)^2}
\frac{dh(\xi)}{d\xi}.
\end{equation}
Since ordinary solitons in the FGM are off-resonance to the
atomic transition, the effect of the atomic system on their
propagation is always weak, unlike the case of SIT in empty
space.

Next we study the multiparticle excitations of the system
with eigenenergies lying inside the frequency gap. We look
for an image of a Bethe string when the real parts of particle
frequencies $\omega_j$ lies inside the gap,
$\xi\in G=(\Omega_\bot,\Omega_\|)$. To find the analytical
continuations of the functions $k(\omega)$ and $h(\omega)$,
an appropriate branch of the function $n(\omega)$ is fixed
by the condition $n(\xi\pm i0)=\pm i\nu(\xi)$, where
$\nu(\xi)=\sqrt{|\varepsilon(\xi)|}$. Soliton parameters
$\xi_j$ and $\eta_j$ are determined by the equations
\end{mathletters}
\begin{equation}
{\rm Re}\,h(\xi_j,\eta_j)=H,\;\;\;{\rm Im}\,h(\xi_j,\eta_j)=\beta(l+1/2-j).
\end{equation}
Approximate solutions of eqs. (7) are easily found by the
method of Ref. \cite{JR}. One can show that in the model
under consideration gap excitations exist only for
$\xi\in(\omega_{12},\Omega_\|)$. A string with an even
number of particles, $n=2l$, describes a bound complex of
$l$ pairs of confined gap excitations - quantum gap soliton -
with the eigenenergy per pair of
\begin{equation}
\epsilon_l=\epsilon^{(0)}_l-\Delta_l+q^2/2m_l,
\end{equation}
where $q$ is the soliton momentum per pair, while
$\Delta_l=\frac{b\beta^2}{12a^3}(4l^2-1)$ is the band width,
and
$m_l=\frac{a}{2b}\left(\frac{1}{l}\sum_{j}|\kappa'(\xi^{(0)}_j)|\right)^2$
is the mass of a gap soliton containing $l$ pairs. Here
$\kappa'=d\kappa/d\xi$, $\kappa(\xi)=\xi\nu(\xi)$, and
$\xi^{(0)}_j=\omega_{12}+(\beta/a)(l+1/2-j)$. The parameters
$a$ and $b$ are found as coefficients in the Taylor series of
the function $\phi(\xi)=[\xi\nu^3(\xi)]^{-1}$ at the point
$\xi=\omega_{12}$: $\phi(\xi)\simeq a+b(\xi-\omega_{12})$.
The positions of the centers of soliton bands are given by
$\epsilon_l^{(0)}=l^{-1}\sum_{j}\xi^{(0)}_j=\omega_{12}+(\beta/a)l$.

The spatial size of a pair, $\delta\simeq\kappa^{-1}(\xi)$,
is nothing but the penetration length of the radiation with
the frequency $\omega=\xi\in G$ into the medium, and hence
it lies on the scale of a few wavelengths. Since $\kappa(\xi)$ is
monotonically decreasing function, the gap soliton size, in
a sharp contrast to the case of ordinary solitons, grows
with the growth of the number of pairs $l$.

For a gap soliton, the dispersion relation inside the atomic
system takes the form
\begin{mathletters}
\begin{equation}
Q(\epsilon_l)=q(\epsilon_l)-\frac{\rho}{l}\arctan{\frac{\beta l}{H}},
\end{equation}
The group velocities of gap soliton propagation inside,
$V_l=d\epsilon_l/dQ$, and outside, $v_l=d\epsilon/dq$,
the atomic system are then related by
\begin{equation}
\frac{1}{V_l}=\frac{1}{v_l}\left(1-
\frac{al}{\sum_{j=1}^{l}|\kappa'(\xi^{(0)}_j)|}\,
\frac{\rho\beta}{H^2+\beta^2l^2}\right).
\end{equation}
Again in a sharp contrast to the case of ordinary solitons,
the velocity of gap soliton propagation inside the atomic
system is greater than $v_l$, i. e., the particle-atom
scattering speeds up a gap soliton. But, it should be emphasized
that the results obtained are valid only for quite small
soliton momenta $q$ when the gap soliton dispersion is
described in the effective mass approximation. At arbitrary
$q$ or for quite big solitons ($l\gg 1$), we cannot use only
the first terms of the Taylor expansion for the function
$\phi(\xi)$ and have to solve the exact equations (7).

So far, we have looked for a soliton image of a string
assuming that all of the string rapidities are mapped to
the soliton frequencies whose real parts lie either outside
(ordinary soliton) or inside the gap (gap soliton). It is
easy to see that this assumption is not necessary, and one
can construct bound many-particle complexes - ``composite''
solitons - containing excitations from different frequency
intervals $C_-$ and $G$. To clarify this point, let us
consider first the simplest example of a three-particle
string with a negative carrying rapidity $H$. Then, the
pair of the complex conjugated rapidities $h_1=H+i\beta$
and $h_3=h^*_1=H-i\beta$ is mapped to the pair of frequencies
$\omega=\xi+i\eta$ and $\omega^*=\xi-i\eta$, where
$\xi=\omega_{12}+\beta/a\in G$ and $\eta=|H|/a$, and momenta
$k=q+i\kappa(\xi)$ and $k^*=q-i\kappa(\xi)$, where
$q=\eta|\kappa'(\xi)|$. Under the condition $H<0$, the real
rapidity $h_2=H$ can be mapped to the real frequency
$\xi_-\in C_-$ and the corresponding real momentum
$k_-=k(\xi_-)$, where $\xi_-$ is determined by $h(\xi_-)=H$.
All three particles compose a bound state of a pair of gap
excitations and a polariton of the lower branch.

The generalization of the composite soliton construction
to the many-particle case is straightforward. Consider an
$N$-particle string with an negative $H$. A pair of the
complex conjugated rapidities or any number of such pairs
could be mapped to corresponding pairs of frequencies whose
real parts lie inside the gap, while the rest rapidities
of the string are mapped to frequencies whose real parts
lie below the gap. In other words, one part of the string
rapidities is used to construct a ``gap soliton'', while
the rest rapidities of the same string are used to construct
an ``ordinary soliton''. All the particles together compose
a bound many-particle complex, so that an composite soliton
can be treated as a bound state of an ``ordinary'' and a
``gap soliton'', but these ``solitons'' are different from
ordinary and gap solitons discussed above and do not exist
separately from each other.

Both gap and composite solitons realize the self-induced
transparency effect in FGM. These solitons describe an entirely
new mechanism for energy transfer in a frequency gap medium.
The results obtained illustrate the rich variety of nonclassical
optical propagation effects within a classically forbidden
frequency gap. Furthermore, they suggest that a doped FGM
with suitable optical pumping may act as a source of novel
quantum correlated states of light.

V. R. is grateful to the Department of Physics at the University
of Toronto for kind hospitality and support. This work was
supported in part by NSERC of Canada and the Ontario Laser
and Lightwave Research Centre.

\end{mathletters}

\end{document}